\documentclass[twocolumn,longauthor,%
breaklinks,colorlinks,urlcolor=blue,citecolor=blue,linkcolor=blue]%
{aastex62}
\usepackage{times}
\journalinfo{The Astrophysical Journal, 2018, in press}

\shorttitle{Reconnection-Generated Coronal Alfv\'{e}n Waves}
\shortauthors{S.\  R.\  Cranmer}

\begin{document}

\title{\bf Low-frequency Alfv\'{e}n Waves Produced by Magnetic
Reconnection in the Sun's Magnetic Carpet}

\author[0000-0002-3699-3134]{Steven R. Cranmer}
\affil{Department of Astrophysical and Planetary Sciences,
Laboratory for Atmospheric and Space Physics,
University of Colorado, Boulder, CO 80309, USA}

\begin{abstract}
The solar corona is a hot, dynamic, and highly magnetized plasma
environment whose source of energy is not yet well understood.
One leading contender for that energy source is the dissipation of
magnetohydrodynamic (MHD) waves or turbulent fluctuations.
Many wave-heating models for the corona and the solar wind presume that
these fluctuations originate at or below the Sun's photosphere.
However, this paper investigates the idea that magnetic reconnection
may generate an additional source of MHD waves over a gradual range of
heights in the low corona.
A time-dependent Monte Carlo simulation of the mixed-polarity magnetic
field is used to predict the properties of reconnection-driven coronal
MHD waves.
The total power in these waves is typically small in comparison to
that of photosphere-driven waves, but their frequencies are much lower.
Reconnection-driven waves begin to dominate the total power spectrum
at periods longer than about 30 minutes.
Thus, they may need to be taken into account in order to understand the
low-frequency power-law spectra observed by both coronal spectropolarimetry
and in~situ particle/field instruments.
These low-frequency Alfv\'{e}n waves should carry more magnetic energy
than kinetic energy, and thus they may produce less nonthermal Doppler
broadening (in comparison to photosphere-driven high-frequency waves)
in emission lines observed above the solar limb.
\end{abstract}


\section{Introduction}
\label{sec:intro}

The physical processes responsible for heating the solar corona and
accelerating the solar wind have not yet been identified definitively.
Some theoretical proposals involve magnetohydrodynamic (MHD) waves
that originate at or below the solar photosphere, propagate up into
the corona, and dissipate their energy as heat
\citep[e.g.,][]{Al47,Os61,Ho86,Mt99,SI06,CvB07,Li14}.
Other ideas focus on the energy released by magnetic reconnection
events in the upper atmosphere \citep{G64,P72,P88,HP84,PD12}
that may break open field lines that were formerly closed
\citep{Fi99,Fi03,Fi05,An11,Ed12}.
Both classes of models appear to predict heating that is highly
intermittent in both space and time \citep[e.g.,][]{vB11,Kl15}
as is observed in the corona \citep{Fl15}.
This fact---combined with the existence of a highly structured
``magnetic carpet'' of mixed-polarity fields at the coronal base
\citep{TS98}---has led to the conjecture that a realistic description
of coronal heating involves {\em both} waves and reconnection.

It should come as no surprise that oscillatory MHD fluctuations
(i.e., waves, shocks, and turbulent eddies) and magnetic reconnection
events coexist with one another and may be of comparable importance
in the corona's energy budget.
In fact, it is increasingly difficult to find simulations that contain
only turbulence with no reconnection, or only reconnection with no
turbulence \citep[see, e.g.,][]{Ve15}.
The focus of this paper is on one aspect of this linkage:
the spontaneous generation of MHD waves from discrete magnetic
reconnection events in the corona.
This general idea has been studied in the past from the standpoint of
infrequent flux-cancellation events \citep{Ho90} or a persistent
``furnace'' of mixed-polarity activity in the supergranular
network \citep{Ax92,RB98}.
Increasingly sophisticated models of reconnection in both the
chromosphere \citep{St99,Is08} and the corona \citep{Ly14,Th17,Kp17,Tr17}
also show wavelike oscillations as a natural by-product.

Coronal jets and solar flares are notable and observable testbeds for
understanding how reconnection produces waves.
Jets are dense and collimated eruptions that appear to accelerate plasma
into open-field regions connected to the solar wind \citep{Rao16}.
These relatively infrequent events often are seen to produce
swaying motions of the magnetic field, indicative of transverse or
toroidal MHD waves \citep{Ci07,Mo11,Mo15,Ya15,Sz17}.
Simulations also predict the generation of waves and turbulence from
individual jets \citep{Le15,Je15,Wy16,Ur17}.
Solar flares are powered by intense bursts of magnetic reconnection,
and they may generate downward-propagating MHD waves that are important
to the subsequent radiative emission \citep{ES82,FH08,RR16}.
Oscillatory behavior has also been seen above flare loop-tops \citep{TS16},
in flare ribbons \citep{Br15}, and in post-flare arcades \citep{Vw05}.

This paper represents an attempt to predict some broad properties
(e.g., energy levels and power spectra) of the MHD fluctuations
generated by the Sun's magnetic carpet.
Section \ref{sec:monte} describes the Monte Carlo simulations that were
used to model the relevant reconnection and loop-opening (RLO) processes.
Section \ref{sec:waves:method} outlines a method of extracting the
fluxes and magnetic perturbation profiles of Alfv\'{e}nic pulses
associated with individual reconnection events in the simulation.
Section \ref{sec:waves:results} then presents simulated wave power
spectra that were extracted from the simulations, and discusses
time-averaged energy densities.
Section \ref{sec:nonwkb} contains some speculation about how these
waves are likely to propagate up through the corona in a manner that
departs from the standard WKB (Wentzel, Kramers, Brillouin) theory.
Lastly, Section \ref{sec:conc} gives a brief summary of the major
results and discusses some of the broader implications of this work.

\section{The Monte Carlo Magnetic Carpet Model}
\label{sec:monte}

\begin{figure}
\epsscale{1.10}
\plotone{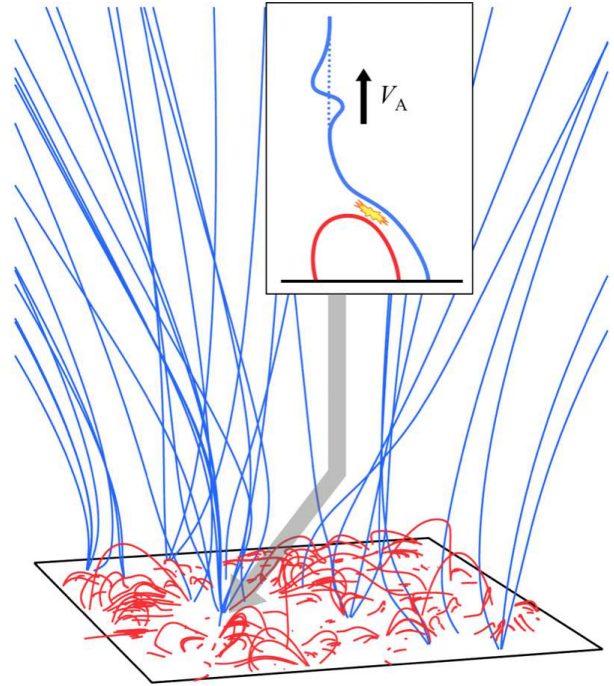}
\caption{Open (blue) and closed (red) magnetic field lines
traced in three-dimensional space for an example timestep 
in the Monte Carlo model with $\xi = 0.5$ and absolute
flux density $B_{\rm abs} = 4$ Mx~cm$^{-2}$.
The horizontal box outlines the (200 Mm)$^2$ photospheric
simulation domain at the base, and the maximum height of the
closed field lines is 18 Mm.
The inset shows a cartoon illustration of a single ``interchange
reconnection'' event that produces an upward propagating
wavelike perturbation along a newly opened field line.
\label{fig01}}
\end{figure}

\citet{CvB10} developed a three-dimensional simulation of
photospheric magnetic flux transport that was coupled with a
potential-field coronal extrapolation.
The main goal of this model was to determine the rates at which closed
field lines open up (i.e., to find the recycling timescale for open
flux) and to estimate how much magnetic energy is released
in reconnection events that involve the opening up of closed loops.
Figure \ref{fig01} illustrates the closed and open field lines
associated with a typical time snapshot of the model.

The \citet{CvB10} simulations comprise four successive stages:
\begin{enumerate}
\item
The complex photospheric field is modeled by means of
a Monte Carlo ensemble of positive and negative monopole sources of
magnetic flux \citep[see, e.g.,][]{Sj97}.
The pointlike flux sources are
assumed to emerge from below (in bipolar pairs), randomly diffuse
across the surface, merge or cancel with their neighbors, and
occasionally fragment into multiple pieces.
The system is evolved with discrete timesteps $\Delta t$,
and it approaches a state of dynamical equilibrium in which
the net rate of magnetic cancellation balances the rate of emergence.
\item
At each new timestep, the coronal magnetic field is recomputed
from the current configuration of photospheric flux elements.
Each element is assumed to act as a monopole-type source, and the
summed vector extrapolation is a so-called {\em potential field.}
Strictly speaking, this is a minimum-energy state that is unable to
release ``free energy'' by undergoing magnetic reconnection.
The relative shortcomings and advantages of this idealization were 
discussed in detail by \citet{CvB10}, but it is important to note that
the potential-field assumption has been found to be useful in at least
identifying the locations of the small regions in which coronal
reconnection must occur \citep{Cl03,Cl05,Rg08}.
\item
Given a range of criteria that locate photospheric flux elements that
``survive'' from one timestep to the next, the \citet{CvB10} code
searches for events in which an initially closed field line at time
$t$ changes into an open field line at $t + \Delta t$.
These are assumed to be the sites of interchange reconnection.
For each of these events, the code outputs the magnitude of magnetic
flux $\Phi_{\rm co}$ connected to the footpoint that remains rooted
to the surface, the horizontal distance $d$ to the other footpoint
at time $t$, and the mean $x,y$ position of the footpoint (averaged
over time steps $t$ and $t + \Delta t$).
\item
Lastly, the amount of nonpotential energy release at each
coronal reconnection event is estimated using a quasi-static
minimum-current corona (MCC) theory \citep{Lg96}.
This model depends on a time-averaged idealization of the amount of
current that must build up---and subsequently dissipate---along a
magnetic separator in the corona.
It does not specify how the dissipated free energy is partitioned
into other forms such as thermal energy, bulk kinetic energy, waves,
and energetic particles.
However, the MCC theory has been found to provide realistic predictions
of the overall reconnection rate \citep[see also][]{Lg01,LK99,BL06,TL12}.
In the present simulations, the total rate of energy loss (summed
over the domain shown in Figure \ref{fig01}) is found to be
appropriately bursty and nanoflare-like.
\end{enumerate}
As a shorthand notation, the phrase ``Monte Carlo model''
will be used in this paper to describe the entire \citet{CvB10}
simulation, despite the fact that only the first of the above four
stages depends on actual Monte Carlo randomization.

\citet{CvB10} summarized the MCC method of estimating the power
$P_{\rm co}$ released by each reconnection event, with
\begin{equation}
  P_{\rm co} \, = \, \theta_{\rm L} C_{\rm L}
  \frac{\Phi_{\rm co}}{d} \left| \frac{d\Phi}{dt} \right| \,\, ,
\end{equation}
and $\theta_{\rm L}$ and $C_{\rm L}$ are dimensionless constants
that describe the efficiency and magnetic separator geometry.
It is assumed that the Monte Carlo time step $\Delta t$ characterizes
the timescale over which individual reconnection events take place,
so the time derivative above can be estimated as
$|d\Phi / dt| \approx \Phi_{\rm co} / \Delta t$.
Thus, the total free energy $E_{\rm co}$ released by each event is
given by the power multiplied by $\Delta t$, and
\begin{equation}
  E_{\rm co} \, \approx \, \theta_{\rm L} C_{\rm L}
  \frac{\Phi_{\rm co}^2}{d} \,\, .
  \label{eq:Eco}
\end{equation}
This quantity is used below to estimate the maximum energy available
to drive an Alfv\'{e}n-wave packet associated with a given event.
The value of the discrete timestep was chosen to be
$\Delta t = 5$~min.
Larger values were found to sometimes skip over some potentially
important energy-release events.
Smaller values of $\Delta t$ ended up subdividing some reconnection
events into smaller pieces, but with summed values of $E_{\rm co}$
being similar to the standard choice of 5~min.
However, with such short timesteps (e.g., $\Delta t \lesssim 1$~min),
the photospheric spatial dimensions traversed are on granular scales.
The \citet{CvB10} Monte Carlo model treats the motions of flux elements
on these scales as essentially random-walk diffusion.
Thus, timesteps shorter than the ones used here would only be appropriate
if the coherent granular motions were being modeled explicitly.

\citet{CvB10} computed the total time-averaged energy flux
associated with the loop-opening events described above.
The resulting values depended strongly on the local value of the
magnetic flux imbalance fraction $\xi$ \citep[see, e.g.,][]{WS04,Hg08}.
This quantity is defined as the ratio of the net magnetic flux density
$B_{\rm net}$ to the absolute unsigned flux density $B_{\rm abs}$,
where
\begin{equation}
  B_{\rm net} \, = \, |B_{+} + B_{-}|  \,\, , \,\,\,\,\,\,\,
  B_{\rm abs} \, = \, B_{+} + |B_{-}|  \,\, ,
\end{equation}
and $B_{+}$ and $B_{-}$ are the mean magnetic flux densities in
regions having positive and negative polarity, respectively.
Quiet-Sun regions with balanced fluxes ($\xi \lesssim 0.5$) were
found to generate far too little energy compared to that required
to heat or accelerate the source regions of the slow solar wind.
Coronal-hole regions with imbalanced fluxes ($\xi \approx 1$)
had larger reconnection-generated energy fluxes, but their recycling
times were much longer than the time it takes the fast solar wind
to accelerate into the low corona.
Thus, the primary conclusion of \citet{CvB10} was that RLO
energy-release processes probably are not responsible for
the majority of either the fast or slow solar wind.
Similar conclusions have been found by others
\citep{KP11,Li16}, but the intermittent jetlike fluctuations
generated by RLO processes are likely to be important in other ways.

Although the original Monte Carlo model successfully predicted a number
of observed properties of the photosphere and corona, it made some key
simplifying assumptions that ought to be acknowledged and reviewed.
\begin{enumerate}
\item
\citet{CvB10} assumed that the corona evolves from one potential-field
state to another, and that the MCC model accurately predicts how much
non-potential energy is released by magnetic reconnection events.
The actual solar field always has a nonpotential
component \citep[e.g.,][]{Pv97,Ye10} that ought to be computed
and not just estimated.
Improved models of the magnetic carpet \citep{My13,Wi15} and
interchange reconnection near open-field regions \citep{Ed12,Hi17}
illustrate the buildup of nonpotential fields on the spatial and
time scales studied in this paper.
\item
The photospheric magnetic parameters of the Monte Carlo model were
based on observational data available in 2010.
In recent years, however, there have been several new high-resolution
observations that could be used to refine these parameters.
For example, the Imaging Magnetograph eXperiment (IMaX) that flew on
the balloon-borne {\em Sunrise} observatory \citep{MP11} was used to
infer magnetic recycling times as short as 12 minutes \citep{Wi13}.
This is at least a factor of 5 shorter than the recycling
timescales seen in the \citet{CvB10} simulations.
Also, IMaX magnetograms were found to resolve photospheric features
with fluxes as small as $9 \times 10^{14}$~Mx \citep{An17}, which
is about a factor of 16 smaller than the resolution limit of the
Monte Carlo model.
It remains to be seen whether these tiny bipolar regions have a
noticeable effect on the supergranular-scale structure and dynamics
of the corona.
\item
The \citet{CvB10} models did not contain explicit supergranular flows,
but they did show a spontaneous aggregation of magnetic fields on
spatial scales of order 15--30 Mm.
This network-like behavior seemed to arise naturally from the
combined action of smaller-scale flux emergence, diffusion, and
cancellation events.
Although this result can be considered evidence for a non-convective
origin for the Sun's supergranulation \citep[see also][]{Ra03,Ch07},
this is still an open question \citep{CR16,FH16}.
If Monte Carlo flux-transport simulations are improved---i.e., by
modeling the nonpotential field or including newly observed
small features---they should be studied to see if this emergent
supergranular aggregation effect survives.
\end{enumerate}
In future work, the issues listed above should be explored, either by
more rigorous testing of existing assumptions or by improving the
physics.
For now, however, the original \citet{CvB10} model will be used to
predict the properties of wavelike fluctuations generated by
magnetic reconnection.

\section{Reconnection-Driven Wave Generation}
\label{sec:waves}

\subsection{Methodology}
\label{sec:waves:method}

The basic assumption of this paper is that each reconnection event
generates a compact ``packet'' in which there is a single transverse
perturbation in the background magnetic field.
For the Cartesian geometry shown in Figure \ref{fig01},
the background field at large heights points upwards, parallel to
the $z$ axis, and the perturbation also propagates along $z$.
(The horizontal spread of the open field seen in Figure \ref{fig01}
is a consequence of modeling only a finite patch of the solar surface
and not its surroundings.)
Each packet is modeled with cylindrical symmetry around a vertically
oriented field-line axis, and its geometric center
has Cartesian coordinates $(x_0, y_0, z_0)$.
The packet flows upward at the Alfv\'{e}n speed
$V_{\rm A} = B/ \sqrt{4\pi\rho}$, where $B$ and $\rho$ are
representative values of the background coronal field strength and
mass density, respectively.
At a sufficient distance above the reconnection region, the time
variability due to the perturbed magnetic field comes solely from
the upward drift of the packet's center, with $z_0 = V_{\rm A} t$.
The coordinates $x_0$ and $y_0$ are those of the mean footpoint
location as output from the Monte Carlo code.
The spatial extent of the magnetic perturbation is assumed to have
a compact Gaussian form, and
\begin{equation}
  B_{\perp} (x,y,z) \, = \, B_{\perp 0} \,
  \exp \left( - \frac{r^2}{\sigma_r^2} \right)
  \exp \left[ - \left( \frac{z - z_0}{\sigma_z} \right)^2 \right]
  \,\, ,
  \label{eq:Bgauss}
\end{equation}
where
\begin{equation}
  r^2 \, = \, (x - x_0)^2 \, + \, (y - y_0)^2 \,\, ,
\end{equation}
and the parameters $\sigma_r$ and $\sigma_z$ describe the spatial
extent of the packet.
Assuming the packet's transverse perturbation has a linear
polarization, its azimuthal orientation is specified by an
angle $\phi$, such that
\begin{equation}
  B_x \, = \, B_{\perp} \cos\phi
  \,\, , \,\,\,\,\,\,\,
  B_y \, = \, B_{\perp} \sin\phi \,\, .
\end{equation}
The total perturbed magnetic energy in the packet $E_{\rm B}$
is given by integrating over the spatial volume, with
\begin{equation}
  E_{\rm B} \, = \, \int dV \, \frac{B_{\perp}^2}{8\pi}
  \, = \, \sqrt{\frac{\pi}{2}} \left(
  \frac{B_{\perp 0}^2 \sigma_z \sigma_r^2}{16} \right)
  \,\, .
  \label{eq:EB}
\end{equation}
If the energy quantity $E_{\rm B}$ and the size parameters
$\sigma_r$ and $\sigma_z$ are known, then the magnetic perturbation
amplitude $B_{\perp 0}$ is determined uniquely.

The free energy released by magnetic reconnection is
partitioned into multiple forms such as thermal heating, nonthermal
particle acceleration, bulk MHD flows, tearing-mode eddies, and waves.
MHD simulations have been helpful in predicting the relative
distribution of energy into these different forms
\citep[see, e.g.,][]{Bi09,Ki10,LT12,Li17}.
Kinetic models reveal the possibility of many additional paths
for energy transfer via instabilities and collisionless effects
\citep{MS92,Da11,Sh11,Fj14,HH15}.

For the purpose of this paper, all we really need to know is
the fraction of the total released free energy $E_{\rm co}$ that
goes into a given packet's transverse field-line perturbation.
Because reconnection simulations are still not clear on how to
compute this fraction (which we will call $f_{\rm B}$), we will
just estimate a likely value for it and study how the subsequent
results are sensitive to how it may vary.
Thus, Equation (\ref{eq:Eco}) is used to estimate
$E_{\rm B} = f_{\rm B} E_{\rm co}$ for each event, and then
Equation (\ref{eq:EB}) is solved for
\begin{equation}
  B_{\perp 0} \, \approx \,
  3.57 \sqrt{\frac{f_{\rm B} E_{\rm co}}{\sigma_z \sigma_r^2}} \,\, .
  \label{eq:Bperp0}
\end{equation}
Lastly, the cylindrical packet-size parameters $\sigma_z$ and
$\sigma_r$ can be estimated as follows.
The Monte Carlo timestep $\Delta t$ was described above
as the time over which a given reconnection event occurs.
Thus, given the assumption of passive advection of the packet at
speed $V_{\rm A}$, it is reasonable to assume that
\begin{equation}
  \sigma_z \, \approx \, V_{\rm A} \Delta t \,\, .
  \label{eq:sigmaz}
\end{equation}
The cylindrical end-cap surface area $\pi \sigma_r^2$ is related
to the degree of horizontal expansion experienced by a bundle of
field lines that originate at the closed footpoint (subtending
magnetic flux $\Phi_{\rm co}$) and extend up into the corona.
\citet{CvB10} described how each Monte Carlo simulation evolves
to a dynamical steady-state with a given mean magnetic flux
density.
At heights above the closed loop-tops, it is the net flux density
$B_{\rm net}$ that essentially prescribes the average strength of
the open field.
Thus, magnetic flux conservation---for the field lines connected
to a given loop opening event---demands that
\begin{equation}
  \Phi_{\rm co} \, = \, \pi \sigma_r^2 \, B_{\rm net} \,\, ,
  \label{eq:fluxcon}
\end{equation}
which is solved for $\sigma_r$ using a mean value of
$B_{\rm net}$ (averaged over the entire simulation in space and time)
and each event's tabulated value of $\Phi_{\rm co}$.

The assumption of a single transverse magnetic pulse in
Equation (\ref{eq:Bgauss}) is likely to be an oversimplification.
MHD simulations \citep[e.g.,][]{Ly14} often show some higher-frequency
oscillations in the reconnection outflow region due to the spontaneous
growth and evolution of magnetic islands.
Other simulations \citep{Ta17} show complex nonlinear mode conversion
around magnetic nulls in the low corona.
Thus, our assumptions that the pulse (1) perturbs the background field
monotonically and (2) propagates with the linear Alfv\'{e}n speed
$V_{\rm A}$ may need to be reexamined in the future.
Nevertheless, it is worthwhile to explore the quantitative properties
of MHD waves associated with this relatively simple starting point.

In practice, a representative coronal ``wavetrain'' is computed by
choosing a measurement point with $x,y$ coordinates at the center of
the Monte Carlo simulation box, and at a height $z$ above the peaks of
the tallest loops.
The exact value chosen for $z$ is relatively unimportant because
Equation (\ref{eq:fluxcon}) does not contain the Sun's large-scale
spherical (or superradial) magnetic expansion.
The only real consequence of choosing a value for $z$ is that it
determines $V_{\rm A}$ (see below).
In any case, the wave packets, with their own unique values of
$x_0$, $y_0$, and $z_0$, are assumed to propagate up past the
measurement point as a function of time $t$.
Each packet has a nonzero influence on the measurement point via
Equation (\ref{eq:Bgauss}), but the most distant ones have a
negligible impact.
Each packet's transverse orientation angle $\phi$ is sampled from
a uniform random distribution.
Lastly, the summed effect of all wave packets is computed to obtain
$B_x(t)$ and $B_y(t)$ at the measurement point, and their Fourier
transforms provide the magnetic fluctuation power spectrum $P_{\rm B}(f)$.

\subsection{Results from the Monte Carlo Models}
\label{sec:waves:results}

Six new Monte Carlo models were created with flux imbalance fractions
$\xi = 0.1$, 0.3, 0.5, 0.7, 0.9, and 0.99.
All input parameters to the code were identical to those described
by \citet{CvB10}, and each model was run with its own unique
random-number seed.
The photospheric models were evolved for 60 days of simulation time,
with $\Delta t = 5$ minutes.
Coronal field lines were traced only during the final 10 days
of each simluation, in order to safely avoid times during which the
system has not yet settled into a dynamical steady state.
The photospheric simulation domain was assumed to be a square box
200 Mm on a side, and it contained discrete flux elements made
up of integer multiples of $10^{17}$~Mx.
However, the coronal field-line tracing algorithm splits each element
up into 7 distributed pieces, each of which can be either open or
closed.
Numbers given below refer to these subdivided elements that have a
magnetic flux resolution of $1.43 \times 10^{16}$~Mx.

After each simulation is evolved for sufficient time to forget its
initial conditions, the number of flux elements $N_{\rm el}$
begins to fluctuate around a fixed mean value.
This mean value is a monotonically decreasing function of $\xi$.
The most magnetically balanced model ($\xi = 0.1$) has
$N_{\rm el} = 1145$, and the most imbalanced model ($\xi = 0.99$)
has $N_{\rm el} = 588$.
On average, the fraction of elements that survive from $t$ to
$t + \Delta t$ is between 0.82 and 0.95.
The fraction of elements that undergo loop-opening is small
(usually of order 0.005 to 0.025), so the number of discrete
closed-to-open events $N_{\rm co}$ that occur in each timestep
is similarly small.
Note, however, that $N_{\rm co}$ is largest for intermediate
values of $\xi$; it increases from a mean value of 10.1 
at $\xi = 0.1$ to a peak of 20.2 at $\xi = 0.5$, then it
decreases to a minimum value of 2.3 at $\xi = 0.99$.
The inherent small-number statistics in these models gives rise to
bursty and intermittent RLO energy release
\citep[see, e.g., Figure 12 of][]{CvB10}.

\begin{figure}
\epsscale{1.15}
\plotone{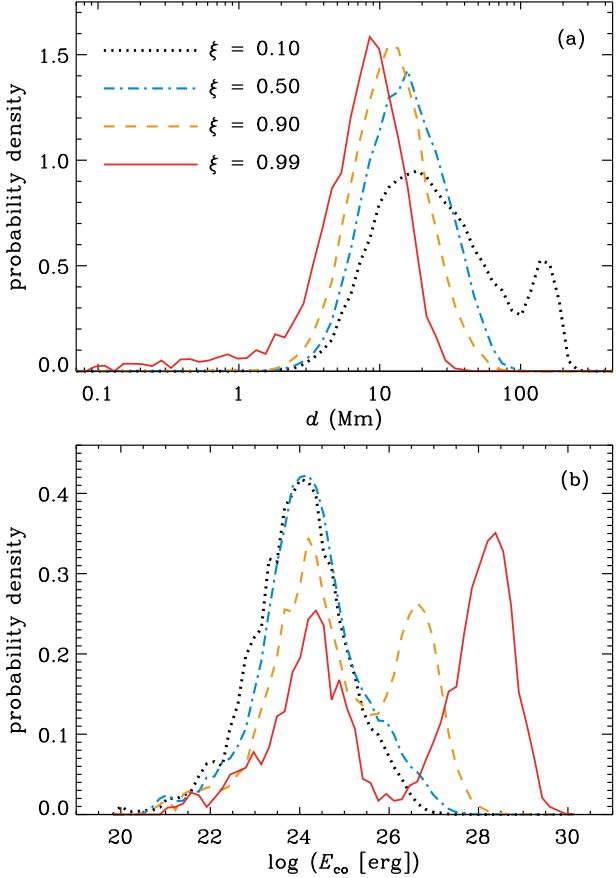}
\caption{Normalized probability distributions for (a) footpoint
separation distance $d$, and (b) estimated free energy $E_{\rm co}$
released by magnetic reconnection in loop-opening events.
Four of the six Monte Carlo models are shown for each quantity, with
$\xi = 0.1$ (black dotted curve),
$\xi = 0.5$ (blue dot-dashed curve),
$\xi = 0.9$ (gold dashed curve),
$\xi = 0.99$ (red solid curve).
The other two models ($\xi = 0.3$, 0.7) have distributions that
fall in between the ones shown.
\label{fig02}}
\end{figure}

Figure \ref{fig02} shows probability distributions for the
footpoint separation distance $d$ and the liberated free energy
$E_{\rm co}$ associated with the database of loop-opening events
output by the Monte Carlo code.
For both quantities, the values were collected into 60 discrete
histogram bins distributed uniformly in the logarithm of either
$d$ or $E_{\rm co}$.
The distributions of $d$ values shown in Figure \ref{fig02}(a)
have median values that decrease monotonically with increasing $\xi$: 
from 22.8 Mm ($\xi = 0.1$) to 7.89 Mm ($\xi = 0.99$).
This corresponds closely to the trend seen in maximum loop heights
(i.e., 95\% percentile values) shown in Figure~7 of \citet{CvB10}.
The mean and median values of loop height fell below these maximum
values, but they followed similar monotonic trends with $\xi$.
Below, we consider $d$ to be an approximate upper limit for the
loop height during an RLO event; i.e., one can safely evaluate the
wave-packet quantities defined in Section \ref{sec:waves:method} at
heights at or above $d$.

The values of $E_{\rm co}$ shown in Figure \ref{fig02}(b) were
computed for a standard assumption of $\theta_{\rm L}C_{\rm L}=0.006$
in Equation (\ref{eq:Eco}).
This is close to the geometric mean of the two limiting values
found by \citet{CvB10} (i.e., 0.003 and 0.011), and it is important
to note that the factor of 4 difference between those numbers is
only a lower limit to the uncertainty range for this quantity.
For most values of $\xi$, the distribution of $E_{\rm co}$ values
is single-peaked, with most-probable values in the nanoflare range
(i.e., $10^{24}$ erg).
However, for cases with $\xi = 0.9$ and $\xi = 0.99$ there is a
marked bimodality, with a second peak reaching up to ``microflare''
energies.
This double-peak structure arises mainly from the distribution
of $\Phi_{\rm co}$ fluxes output by the Monte Carlo code, and is
only very weakly anticorrelated with the distribution of $d$ values.
\citet{CvB10} speculated that these strong ($10^{28}$ erg) events may
correspond to the bright polar jets observed in large coronal holes.

\begin{figure}
\epsscale{1.15}
\plotone{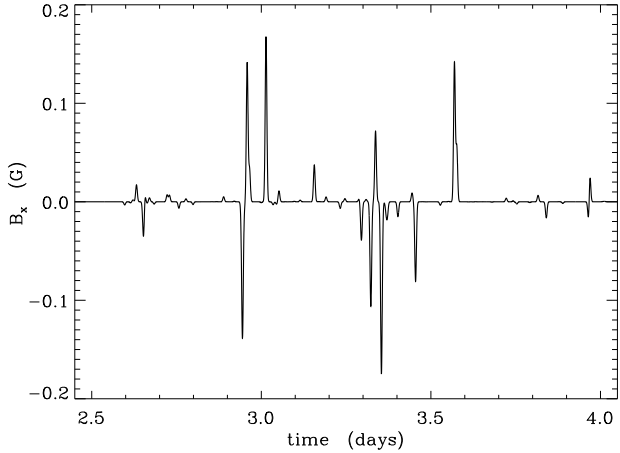}
\caption{Summed magnetic perturbation, due to the full set of
loop-opening events for the $\xi = 0.7$ Monte Carlo model, evaluated
at a measurement height of $z=100$~Mm.
See text for the values of all input parameters.
\label{fig03}}
\end{figure}

Figure \ref{fig03} shows an example $B_x(t)$ waveform for the $\xi = 0.7$
model and a subset of the full 10-day coronal simulation time.
Magnetic perturbations were computed using a fine sampling timestep
of 10 seconds, which resolves individual Gaussian-packet profiles with
30 measurement points per $\Delta t$.
In order to compute these waveforms, values needed to be chosen for
two remaining parameters: $f_{\rm B}$ and $V_{\rm A}$.
For the remainder of this paper, we assume the partition fraction
$f_{\rm B}$ has a value of 0.25.
This is consistent with roughly half of the released free energy going
into an Alfv\'{e}nic pulse that exhibits equipartition
between its kinetic and magnetic energy components \citep{Wa44}.
The Alfv\'{e}n speed $V_{\rm A}$ depends on the height at which
the waves are simulated.
The results presented here assume a representative height $z = 100$~Mm
above the photosphere, which is above the peaks of nearly all loops
but still close enough to the Sun to justify ignoring spherical
expansion effects.
Self-consistent coronal models based on photosphere-driven
waves and turbulence \citep{CvB07,CvB13} give a range of Alfv\'{e}n
speeds at this height between about 1500 and 3200 km~s$^{-1}$,
depending on whether it is a fast or slow solar wind stream.
Thus, we adopted a typical value of $V_{\rm A} = 2000$ km~s$^{-1}$.

\begin{figure}
\epsscale{1.15}
\plotone{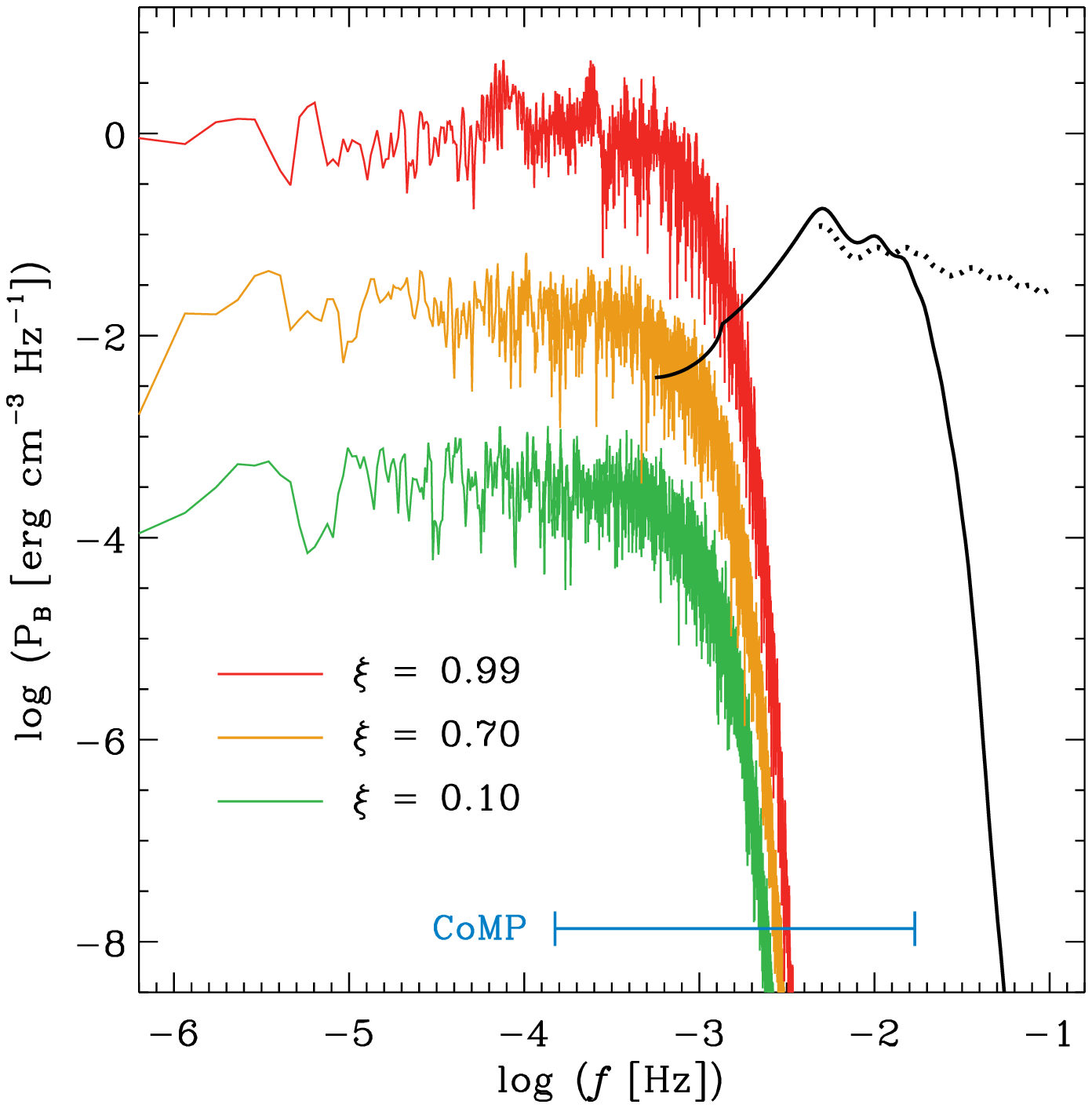}
\caption{Magnetic energy density power spectra for RLO Monte Carlo
models with $\xi = 0.1$ (green curve), $\xi = 0.7$ (gold curve),
and $\xi = 0.99$ (red curve).
Also shown are Alfv\'{e}nic fluctuation spectra driven by photospheric
motions and extrapolated up to $z = 100$~Mm, from the idealized
coronal-hole model of \citet{CvB05} (solid black curve) and
the measurements of \citet{Ch12} (dotted black curve).
The range of frequencies measured by CoMP (horizontal blue bar)
is also noted.
\label{fig04}}
\end{figure}

Stochastic time-series waveforms can often be understood more clearly
when examined in the frequency domain.
For each of the six Monte Carlo models, Fast Fourier Transforms (FFTs)
were performed on the $B_x$ and $B_y$ waveforms, and power spectra were
computed by multiplying each FFT by its own complex conjugate.
Frequency integrals over these power spectra would give the variances
$\langle B_x^2 \rangle$ and $\langle B_y^2 \rangle$.
Lastly, the sum of these two power spectra, divided by $8\pi$,
gives the full transverse magnetic energy power spectrum $P_{\rm B}(f)$.
Thus,
\begin{equation}
  \frac{\langle B_x^2 \rangle}{8\pi} \, + \,
  \frac{\langle B_y^2 \rangle}{8\pi} \, = \,
  \int_0^{\infty} df \,\, P_{\rm B}(f) \, = \, U_{\rm B} \,\, ,
  \label{eq:UB}
\end{equation}
where $U_{\rm B}$ is the time-averaged magnetic energy density due
to transverse fluctuations at the adopted measurement point.
Figure \ref{fig04} shows $P_{\rm B}(f)$ for three of the six
Monte Carlo models.
These spectra tend to be flat (i.e., ``white noise'') for frequencies
less than about $2 \times 10^{-4}$ Hz, and they begin to drop off
exponentially around $10^{-3}$ Hz.
All six models have similar spectral shapes, and they all have
nearly identical values of the most-probable frequency, defined by
\begin{equation}
  \langle f \rangle \, = \,
  \frac{\int df \,\, P_{\rm B}(f) \,\, f}{\int df \,\, P_{\rm B}(f)}
  \, \approx \, 4.27 \times 10^{-4} \,\, \mbox{Hz.}
  \label{eq:fmostprob}
\end{equation}
Among the six models, the standard deviation in $\langle f \rangle$
is only about 3\% of the mean value given above.
The corresponding most-probable period $\langle f \rangle^{-1}$
is approximately 40 minutes.

The computed spectral shapes can be understood as a manifestation
of the assumed pulse shapes shown in Figure \ref{fig03}.
\citet{CvB05} showed that the power spectrum corresponding to a series
of intermittent Gaussian pulses is given by
\begin{equation}
  P_{\rm B} \, \propto \, e^{-2 \pi^2 \tau^2 f^2}
  \label{eq:gaufit}
\end{equation}
where $\tau$ is the $1/e$ half-width of a single pulse
measured at a fixed location in space.
This is essentially the same quantity as the timestep
$\Delta t$ used in Equation (\ref{eq:sigmaz}).
Thus, it is not surprising that the above expression, with
$\tau = 5$~min, is an excellent fit to the frequency dependence
of the curves shown in Figure \ref{fig04}.
In addition, Equation (\ref{eq:gaufit}) provides an exact solution
for the most-probable frequency,
\begin{equation}
  \langle f \rangle \, = \, \frac{1}{\tau \sqrt{2 \pi^3}}
\end{equation}
which also agrees well with the numerical solutions to
Equation (\ref{eq:fmostprob}) discussed above.
Of course, in the real corona, there must be a continuous distribution
of reconnection timescales instead of a single $\Delta t$, so we
expect the resulting power spectra to drop off more gradually with
increasing frequency.

For comparison, Figure \ref{fig04} also shows power spectra that
estimate the coronal magnitudes of MHD waves driven by photospheric
granulation.
The two black curves were derived from horizontal kinetic energy spectra
that in turn were computed from intergranular bright-point motions---one
of them from a semianalytic model based on earlier observations
\citep{CvB05}, and the other from more recent high-resolution data
\citep{Ch12}.
The processing of the latter data was discussed in more detail by
\citet{VK17}, who also found a similar high-frequency power-law tail
in spectra derived from simulations.
These photospheric spectra were extrapolated up to a coronal height
of $z = 100$~Mm and converted to magnetic fluctuation spectra using
the polar coronal-hole model of \citet{CvB05}.

The dominant frequencies of the photospheric waves shown in
Figure \ref{fig04} tend to be much higher than those produced by the
Monte Carlo reconnection model.
However, there is a region of frequency overlap in which both models
may contribute comparably to the total power.
In this region, the Monte Carlo model shows decreasing power as a
function of increasing frequency, and the \citet{CvB05} model shows
increasing power as a function of increasing frequency.
The latter can be understood by examining the propagation history of
these waves from the photosphere to the corona.
Waves near the peak of the black curve ($\log f \approx -2.3$)
are above the photospheric kink-mode cutoff frequency,
so they have been propagating the whole way.\footnote{%
Transverse MHD waves that originate in strong-field
intergranular ``flux tubes'' may become evanescent for frequencies
below a critical cutoff value of $\log f \approx -2.8$ (i.e.,
periods of order 9--12 min).
This gravitational-stratification effect is similar to that
experienced by acoustic waves at a slightly higher cutoff frequency
\citep[see, e.g.,][]{Sp81,HK99}.}
Waves at the low-frequency end of the \citet{CvB05} spectrum
($\log f \approx -3.3$) spent some time below the kink-mode cutoff in
the low chromosphere.
Thus, they experienced some evanescent decay and ended up with
less power in the corona.

The region of frequency overlap between the two sets of models in
Figure \ref{fig04} corresponds to the frequencies measured by the
Coronal Multi-channel Polarimeter \citep[CoMP;][]{Tz08}.
Although CoMP measurements so far do not yet allow us to measure
the absolute wave power in the corona (mainly because of line-of-sight
cancellation of overlapping Doppler signals), they have provided useful
data on the shape of the Alfv\'{e}n-wave power spectrum.
Between frequencies of $10^{-4}$ and $10^{-2}$ Hz, CoMP tends
to show monotonically decreasing power with slopes of order
$P \propto f^{-1}$ to $f^{-1.5}$ \citep{Tz09,Liu14,Mt16}.
Figure \ref{fig04} indicates that an understanding of this
monotonic spectrum may require us to take account of {\em both} the
low-frequency reconnection-driven waves and the high-frequency
photospheric waves.

The remainder of this paper discusses the magnitudes of the
reconnection-generated waves shown in Figure \ref{fig04}.
The magnetic energy density $U_{\rm B}$, as defined in
Equation (\ref{eq:UB}), varies by more than three orders of magnitude
from the
$\xi = 0.1$ model ($U_{\rm B} = 2.7 \times 10^{-7}$ erg cm$^{-3}$)
to the
$\xi = 0.99$ model ($U_{\rm B} = 9.7 \times 10^{-4}$ erg cm$^{-3}$).
This relative increase can be understood by examining the scaling
for the magnetic perturbation due to a single reconnection pulse.
Using Equations (\ref{eq:Eco}), (\ref{eq:Bperp0}), and
(\ref{eq:fluxcon})---and ignoring quantities that remain unchanged
from one value of $\xi$ to another---one can estimate
\begin{equation}
  U_{\rm B} \, \propto \, \frac{\Phi_{\rm co} B_{\rm net}}{d} \,\, .
  \label{eq:UBapprox}
\end{equation}
\citet{CvB10} showed that $B_{\rm net}$ increases by about a factor
of 90 as $\xi$ increases from 0.1 to 0.99.
Note that $B_{\rm net} = \xi B_{\rm abs}$, and $B_{\rm abs}$
increases with $\xi$ because the more unipolar models tend to have
a faster emergence rate of flux elements into the photosphere.
The statistical quantities output from the Monte Carlo model (as
shown in Figure \ref{fig02}) indicate that the mean ratio
$\Phi_{\rm co}/d$ increases by about a factor of 60 
as $\xi$ increases from 0.1 to 0.99.
Thus, Equation (\ref{eq:UBapprox}) predicts a factor of 5400
increase in $U_{\rm B}$ over this range.
This slightly overestimates the factor of 3600 seen in the numerical
models, but Equation (\ref{eq:UBapprox}) is only a simple approximation.
For example, it does not take into account the effect of temporal
intermittency for multiple pulses sampled by a fixed observer.

\begin{figure}
\epsscale{1.15}
\plotone{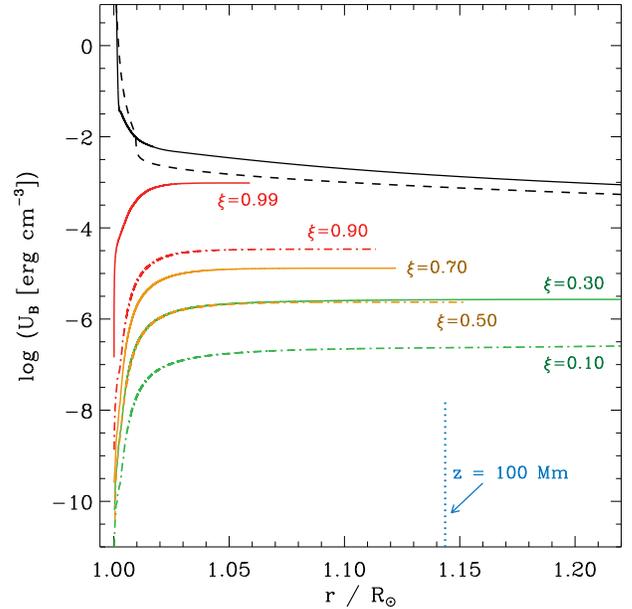}
\caption{Radial dependence of magnetic energy densities $U_{\rm B}$
of coronal waves.
Results from the Monte Carlo models (red, gold, and green curves)
were computed as described in the text.
Models of high-frequency photosphere-generated waves were taken from
\citet{CvB05} (solid black curve) and
\citet{CvB07} (dashed black curve).
\label{fig05}}
\end{figure}

Figure \ref{fig05} shows an estimate of the radial dependence of
each model's wave energy density $U_{\rm B}$.
For the Monte Carlo models, these curves were computed by assuming the
wave energy is built up gradually over the heights corresponding to
each model's distribution of footpoint separation distances $d$.
In other words, a loop-opening event with footpoint separation $d$
is assumed to deposit its free energy as upward-propagating waves only
at heights $z \geq d$.
Thus, the radial functions shown in Figure \ref{fig05} are the
integrated (cumulative) distributions that correspond to the
probability distributions shown in Figure \ref{fig02}(a).
Each curve was then normalized to that model's own value of $U_{\rm B}$.

Figure \ref{fig05} also shows the known radial dependence of magnetic
fluctuation energy density for the photosphere-driven wave models
of \citet{CvB05} and \citet{CvB07}.
This radial dependence is derived from wave-action conservation,
which in the low corona works out to the proportionality
$U_{\rm B} \propto \rho^{1/2}$.
At large distances, the Monte Carlo models also ought to exhibit a
similar radial decline in $U_{\rm B}$ as do the photosphere-driven
waves, but that effect was not included in the multi-color curves
shown in Figure \ref{fig05}.
All of the modeled reconnection-driven waves (except the extreme
case $\xi = 0.99$) have energy densities substantially
weaker than those corresponding to the higher-frequency MHD waves
expected to come from the photosphere.

\section{Non-WKB Properties of the Waves}
\label{sec:nonwkb}

Although MHD waves associated with magnetic-carpet reconnection do
not appear to dominate the total wave energy density in the corona,
they may be responsible for ``filling in'' the lowest frequencies
of the power spectrum.
In interplanetary space, the highest power levels occur at the
lowest frequencies \citep[see, e.g.,][]{TM95,BC13}.
It is still not known what fraction of these low-frequency fluctuations
originates in the corona (or lower), what fraction is produced by
some kind of inverse cascade from the high-frequency turbulence,
and what fraction may be the result of corotating (but otherwise
time-steady) flux tubes advecting past the spacecraft.

Remote-sensing observations can be used to put constraints on the
properties of the coronal MHD-wave spectrum.
Emission-line spectroscopy provides information about
long-time averages of transverse velocity fluctuations
via the so-called ``nonthermal'' component of the line profile
\citep{Bo73,Ma79}.
This information is complementary to the short-time Doppler
fluctuations measured by, e.g., CoMP, as discussed above.
Measurements made above coronal holes over the past few decades
\citep{Tu98,Bj98,CvB05,DS08,LC09} appeared to agree well
with theoretical predictions of undamped Alfv\'{e}n waves launched
at the solar surface.
However, more recent data from the Extreme-ultraviolet Imaging
Spectrometer (EIS) on {\em Hinode} seem to show substantial
wave damping above heights of roughly 0.2 $R_{\odot}$
\citep{Hh12,BA12,Hh13,Gu17}.

There is still no universally agreed-upon explanation for the
apparent wave damping inferred from EIS observations.
Undamped WKB Alfv\'{e}n waves require a radial increase in their
transverse velocity amplitude ($v_{\perp} \propto \rho^{-1/4}$),
but the data show nonthermal line-widths flattening out and possibly
starting to decrease with increasing height.
This may be indicative of some actual wave dissipation \citep{Zh15},
but it also disagrees with earlier models that predicted much
weaker damping in the corona \citep[see, e.g.,][]{Cr17}.
It also disagrees with earlier measurements of larger nonthermal line
widths at heights just barely above those probed by EIS \citep{Es99}.
Earlier studies ruled out contamination by instrumental stray
light---which would add a narrower component to the broad coronal
emission line---but more recent work indicates EIS sometimes
sees a stray-light signal several times stronger than was assumed
previously \citep{WL18}.
Also, the traditional assumption that emission-line profiles are
dominated by motions in the ``plane of the sky'' may not be valid
for all ions.
A given ion's transition from ionization equilibrium at low
heights to frozen-in ionization at large heights needs to be modeled
self-consistently in order to determine the regions that dominate
the observed line profile \citep[see, e.g.,][]{GC18}.

This paper provides another possible explanation for the observational
data: departures from \citet{Wa44} energy equipartition.
This would be a natural consequence of the extremely low-frequency
spectrum associated with reconnection-driven waves.
High-frequency MHD fluctuations are expected to have wavelengths
smaller than the scales of radial variation in the corona.
Thus, they should behave as ideal WKB waves in a homogeneous background,
with equal magnetic and kinetic energy densities
($U_{\rm B} = U_{\rm K}$).
On the other hand, studies of non-WKB wave propagation
\citep[e.g.,][]{HO80,Ba91,MC94}
tend to show how low-frequency (large-wavelength) waves become
reflected by the radial variations and exhibit
$U_{\rm B} > U_{\rm K}$.
It is worthwhile to note that the MHD simulations of \citet{Ly14}
also saw $U_{\rm B} > U_{\rm K}$ for transient fluctuations driven by
reconnection.
With less of the total energy going into kinetic fluctuations, the
observable transverse velocity amplitude $v_{\perp}$ would be
lower than in the WKB limit.

\begin{figure}
\epsscale{1.15}
\plotone{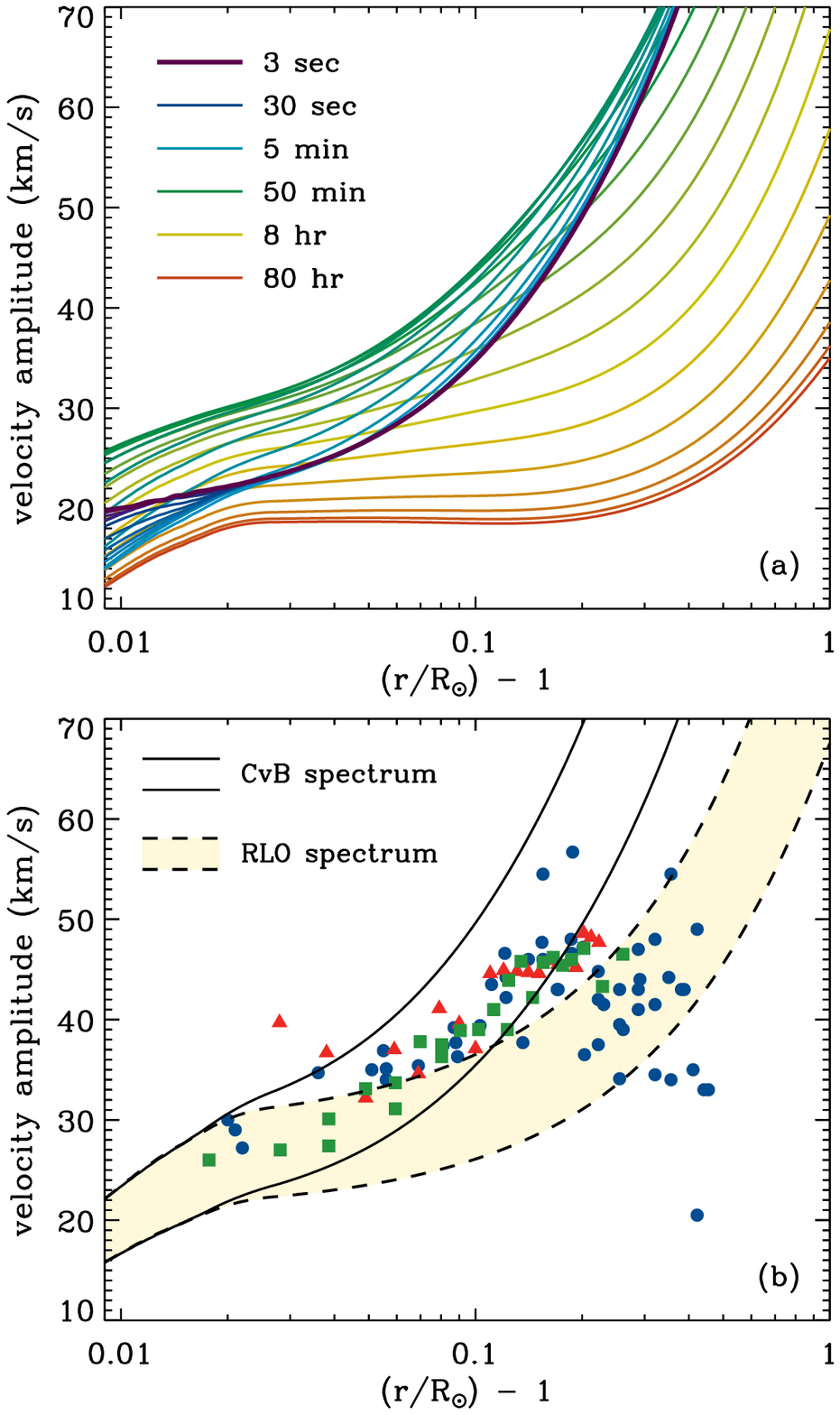}
\caption{Line-of-sight projected Alfv\'{e}n-wave velocity
amplitudes.
(a) Curves computed for 31 discrete wave periods sampled from an
evenly spaced logarithmic grid from 0.05 to 5000 min.
Legend shows one color per decade in period.
(b) Observed nonthermal speeds from
\citet[][green squares]{Bj98}, \citet[][red triangles]{LC09},
and \citet[][blue circles]{Hh13}.
Weighted averages are shown
using power spectra from \citet{CvB05} (solid curves) and
Monte Carlo results of this paper (dashed curves).
\label{fig06}}
\end{figure}

Figure \ref{fig06}(a) shows what the root-mean-squared (rms)
Alfv\'{e}n-wave velocity amplitudes would look like for a range of
monochromatic frequencies.
Each curve is assumed to have the same radial variation of
$U_{\rm tot} = U_{\rm B} + U_{\rm K}$, but the Alfv\'{e}n ratio
$\alpha = U_{\rm K}/U_{\rm B}$ is different for each frequency.
The radial dependences of both
$U_{\rm tot}$ and $\alpha$ are taken from the non-WKB models of
\citet{CvB05}.
Thus, for each curve, the velocity amplitude is computed from
\begin{equation}
  v_{\perp}^2 \, = \,
  \frac{2 U_{\rm K}}{\rho} \, = \,
  \frac{2 U_{\rm tot}}{\rho} \,
  \left( \frac{\alpha}{1 + \alpha} \right) \,\, .
\end{equation}
The plotted velocities were also multiplied by $1/\sqrt{2}$ in order
to show just one projected transverse component.
This allows a more direct comparison with the observed nonthermal
line-widths.
For the shortest periods, $\alpha \approx 1$ and the curves
resemble the classical WKB result ($v_{\perp} \propto \rho^{-1/4}$).
For the longest periods, the curves flatten out in a manner similar
to what is seen in the observations.

Figure \ref{fig06}(b) shows observed coronal-hole data points from 
\citet{Bj98}, \citet{LC09}, and \citet{Hh13}.
The curves indicate weighted-average velocity
amplitudes computed by integrating over a power spectrum, with
\begin{equation}
  \langle v_{\perp}^2 \rangle \, = \, \frac{2}{\rho}
  \int_0^{\infty} df \,\, P_{\rm B}(f) \, \alpha(f) \,\, .
\end{equation}
These curves are presented in pairs, with one multiplied by
$1 / \sqrt{2}$ as discussed above, and the other left alone.
This is meant to illustrate the observed spread in the data, some of
which may be due to intrinsic variability in the wave-generation
regions (e.g., plumes versus interplume regions in coronal holes).
The solid curves were computed from the high-frequency-dominated
photosphere-driven spectrum of \citet{CvB05}, which tends to
resemble the WKB limiting case of $v_{\perp} \propto \rho^{-1/4}$.

The dashed curves in Figure \ref{fig06}(b) were computed using
a somewhat speculative hypothesis that the wave {\em magnitudes}
obey the same radial dependence of $U_{\rm tot}$ used in
Figure \ref{fig06}(a), but the {\em power spectrum} is that of
the reconnection-driven waves as shown in Figure \ref{fig04} and
estimated in Equation (\ref{eq:gaufit}).
Of course, the Monte Carlo model predicts that only a small fraction
of the total power is in the form of these low-frequency
reconnection-driven waves.
However, it is possible that the development of coronal turbulence
drives the spectral shape towards one dominated by the lowest
frequencies.
In this case, the energy partition becomes dominated by the magnetic
fluctuations (i.e., $U_{\rm B} > U_{\rm K}$) and the velocity
amplitude flattens out between heights of 0.05 and 0.30 $R_{\odot}$
in a manner somewhat reminiscent of the EIS data.

\section{Discussion and Conclusions}
\label{sec:conc}

This paper took the output from an existing Monte Carlo model of the
Sun's magnetic carpet \citep{CvB10} and used it to simulate the
properties of low-frequency MHD waves generated by multiple
magnetic reconnection events.
In most regions of mixed magnetic polarity (i.e., everywhere except
the most unipolar regions typified by the $\xi = 0.99$ models), the
total power in reconnection-driven waves is predicted to be much lower
than the power in waves associated with photospheric granulation.
However, the reconnection-driven waves may dominate the coronal
power spectrum at frequencies lower than $10^{-4}$ to $10^{-3}$ Hz.
Thus, obtaining a complete understanding of the turbulent power observed
by off-limb instruments such as CoMP \citep{Tz09} may be predicated on
improving our knowledge about reconnection-generated MHD waves.

The results presented in this paper represent only a cursory survey
of the actual properties of reconnection-driven waves in the corona.
This work needs to be followed by more comprehensive models and more
focused comparisons to the observations.
For example, note that the original \citet{CvB10} Monte Carlo model
was created to address the issue of whether RLO-type events
could be responsible for accelerating the solar wind.
Thus, most of the focus has been on $E_{\rm co}$, the energy in
``closed-to-open'' reconnection events.
\citet{CvB10} found that there also occur ``open-to-closed'' type events
that have mean energies $E_{\rm oc}$ roughly comparable to $E_{\rm co}$.
These events may produce downward-propagating waves, which have been
seen in simulations \citep[e.g.,][]{Ly14} and may be an important
source of counterpropagating wave packets as needed for the production
of a turbulent cascade \citep{Ir63,Kr65,HN13}.

The \citet{CvB10} simulations also did not keep track of
``closed-to-closed'' reconnection events; i.e., those that involve
swapping footpoints between neighboring closed-loop flux systems.
Those kinds of events may dominate the energy budget at low heights
in regions of mixed magnetic polarity.
\citet{Cn16} discussed how a sufficiently complex coronal field-line
topology can trap MHD waves and cause stresses to build up in
magnetically closed regions.
However, if an appreciable fraction of reconnection-driven waves are
in compressible modes that transmit energy {\em across} the field,
then some of that energy may ultimately escape into the solar wind.
In fact, simulations often show that reconnection can give rise to
fast/slow magnetosonic modes \citep[e.g.,][]{Ki10} as well as torsional
Alfv\'{e}nic pulses that resemble flux ropes \citep{HL18}.

Some other limitations of the \citet{CvB10} Monte Carlo model were
listed at the end of Section \ref{sec:monte}.
There is also the sensitivity of the modeled wave power to
the presumed energy partition fraction $f_{\rm B}$, which was only
estimated qualitatively.
Many of these limitations could be addressed by replacing this kind
of model by a fully three-dimensional solution of the MHD conservation
equations \citep[see, e.g.,][]{Am15,Ca16,Rm17,MS17}.
However, an MHD simulation that would encompass the region shown in
Figure \ref{fig01}---and run it for several days of physical
time---remains extremely computationally expensive.
Semi-analytic techniques allow us to simulate the
required domains with both modest resources and more-than-adequate
dynamic range (in space and time).

Lastly, it remains to be seen whether the proposed reconnection-driven
waves are responsible for either: (1) the unexpectedly narrow
{\em Hinode}/EIS line profiles, or (2) the low-frequency-dominated
turbulent power spectra measured in interplanetary space.
For the former, it would be advantageous to repeat the
measurements with a properly occulted coronagraph spectrometer,
such as a next-generation follow-on to the {\em SOHO} Ultraviolet
Coronagraph Spectrometer \citep[see, e.g.,][]{Ko08}.
For the latter, it was mentioned above that much of the low-frequency
variability observed in~situ may be due to the Sun's rotation carrying
uncorrelated bundles of magnetic flux past the spacecraft.
If these fluctuations are similar to larger-scale corotating interaction
regions (CIRs), they may be distinguishable from Alfv\'{e}n waves
from their strong variations in magnetic pressure \citep{CvB13}.
It would also be advantageous to measure solar wind fluctuations from
a spacecraft in strict corotation with the plasma.
{\em Parker Solar Probe} ({\em{PSP}})
will spend a few days near corotation around each perihelion
\citep{Fox16} but probably not long enough to gather sufficient data
to probe the $f < 10^{-4}$~Hz part of the spectrum.

\acknowledgments

\vspace*{0.10in}
The author gratefully acknowledges
Mark Rast, Dana Longcope, Lucas Tarr, and Justin Edmondson
for many valuable discussions.
This work was supported by NSF SHINE program grant AGS-1540094,
NASA Heliophysics Supporting Research grants NNX15AW33G and
NNX16AG87G, and start-up funds from the Department of
Astrophysical and Planetary Sciences at the University of
Colorado Boulder.

\end{document}